\documentclass[a4paper]{article}
\usepackage[pdftex]{graphicx,color}
\usepackage{a4wide}
\usepackage[pdftex]{color}
\usepackage[pdftex,colorlinks=true,bookmarks=false,bookmarksopen=false]{hyperref}

\title{Cosmic-Ray Events as Background in \\
Imaging Atmospheric Cherenkov Telescopes}

\author{G.~Maier$^{1,2}$ and J.~Knapp$^1$ \\[2mm] 
$^1$Department of Physics, McGill University, Montr\'{e}al, Canada H3A 2T8\footnote{
Phone: +1 514 398 6492, email: maierg@physics.mcgill.ca}\\
$^2$School of Physics and Astronomy, University of Leeds, Leeds LS2 9JT, UK}

\begin{document}

\maketitle

\begin{abstract}
The dominant background for observations of $\gamma$-rays in the
energy region above 50 GeV with Imaging Atmospheric Cherenkov 
telescopes  are cosmic-ray events.
The images of most of the cosmic ray showers
look significantly different from those of $\gamma$-rays and are
therefore easily discriminated.  However, a small fraction of events
seems to be indistinguishable from $\gamma$-rays.  This constitutes an
irreducible background to the observation of high-energy $\gamma$-ray
sources, and limits the sensitivity achievable with a given
instrument.  Here, a Monte Carlo study of $\gamma$-like cosmic-ray events
is presented.  The nature of $\gamma$-like cosmic-ray events, the
shower particles that are responsible for the $\gamma$-like appearance,
and the
dependence of these results on the choice of the hadronic interaction
model are investigated.  Most of the $\gamma$-like cosmic ray events are
characterised by the production of high-energy $\pi^0$'s early in the
shower development which dump most of the shower energy into
electromagnetic sub-showers.  Also Cherenkov light from single muons
can mimic $\gamma$-rays in close-by pairs of telescopes.
Differences of up to 25\% in the collection area for $\gamma$-like 
proton showers between QGSJet/FLUKA and Sibyll/FLUKA simulations
have been found.
\end{abstract}

\section{Introduction}

The study of the non-thermal universe at energies above 80 GeV by
means of ground-based $\gamma$-ray astronomy has evolved substantially
in the past few years.  About 40 sources of high energy $\gamma$-rays
are known, including Pulsar Wind Nebula (e.g.~the Crab Nebula),
Supernova Remnants (e.g.~RX J 1713.7-3946), Active Galactic Nuclei
(e.g.~Markarian 421), X-ray binaries (e.g.~LS5039
\cite{Aharonian-2005}), our own galactic centre and even objects that
are not seen in any other waveband (e.g.~HESS J1614-518
\cite{Aharonian-2005b}). This progress is due to the large increase
in sensitivity of new instruments, which consist of arrays of large 
imaging atmospheric Cherenkov telescopes (IACT).  These systems detect
Cherenkov light from air showers simultaneously
in several telescopes. The sensitivity increase is primarily due to
the much improved suppression of the background of hadronic cosmic rays,
which are more than a thousand times more numerous than the $\gamma$-rays.

Arrays of imaging atmospheric Cherenkov telescopes reject a
significant number of cosmic rays already at the trigger level 
by requiring compact patterns (e.g.~more than 3 adjacent pixels) in
two or more telescopes.  
Further suppression is achieved by applying cuts to the shape
parameters describing the images (i.e.~the Cherenkov light
distribution) in the focal plane
For point-like or slightly extended sources
the reconstructed arrival direction can also be used to distinguish
between $\gamma$-rays and the isotropic Cosmic-rays.
The combination of all selection cuts leads to the elimination of the
major part of the background events.
For point-like sources, image
shape and shower direction cuts typically suppress the background by a
factor of 2000. Even so, after all $\gamma$-hadron separation cuts, 
a small but significant fraction of the remaining events 
are of hadronic origin. The large
ratio of cosmic rays to $\gamma$-rays and the substantial
fluctuations in the shower development of hadronic showers lead in
general to a considerable overlap of the distributions of shower parameters, which
are used for the separation. Observations are therefore still
background limited and most of the weaker known sources require observation times in the
range of 10-80 hours for a significant detection.
In general, even
longer times are needed for morphology studies of extended objects,
and for the detection of sources close to the sensitivity limit of the instrument.
Many undetected sources are expected at fluxes just below the current
sensitivity of about $10^{-13}$ cm$^{-2}$ s$^{-1}$ and at energies
between 20 and 250 GeV. Therefore, several groups are currently studying how
to access these regions. Sensitivity improvements and the extension of the
energy range to lower energies requires the collection of more
Cherenkov photons at ground level and improved $\gamma$-hadron separation.  
The former can be achieved, for example, by increasing the size of the telescopes, by
increasing the quantum efficiency of the cameras, or by building more
telescopes.  
The latter relies on profound knowledge of the
development of $\gamma$-ray and cosmic-ray initiated showers in the atmosphere, and requires an
algorithm to identify the subtle differences between them. Both
the improvement of current systems and the design of new observatories are
studied with detailed Monte Carlo simulations of air shower
development and instrument performance.

Unfortunately, it is an immense effort to simulate the background for
current IACTs.
Typically only about one in $10^{6}$-$10^{7}$ simulated proton showers
triggers an array and passes all $\gamma$-hadron separation cuts. This
makes background estimation with simulations, and systematic studies of
telescope designs, trigger conditions, reconstruction procedures
etc.~very time-consuming.

In this paper we study background events in IACTs for two reasons:
firstly, to understand the origin of $\gamma$-like cosmic ray showers
(and perhaps to improve the $\gamma$-hadron separation in the data
analysis), and secondly, to find a feature in the early shower
development in order to distinguish between $\gamma$-like and other proton showers
and to limit the simulation effort to the small fraction of
showers which will finally pass all cuts.

The main characteristics of the simulations, the telescope array and the
analysis method are presented in Section \ref{chap-basics}.  A
detailed study of $\gamma$-like background events is given in Section
\ref{chap-background}.  Section \ref{chap-muons} emphasizes the
importance of the geometrical arrangement of the telescopes and
introduces a
cut against muon-induced background
events for telescope pairs located close to each other. Section
\ref{chap-pi0} describes why some of the background events are
indistinguishable from $\gamma$-ray events and in Section
\ref{chap-models} the differences due to different hadronic
interaction models are discussed.

\section{Simulations and analysis methods}
\label{chap-basics}

Simulations and the subsequent data analysis have been carried out in
several steps.  The first step consists of the simulation of extensive
air showers induced by primary protons and $\gamma$-rays and the
Cherenkov light production by the shower particles.  Next, the
Cherenkov photons are tracked through the telescope optics and the
camera response is modeled.  Finally, the resulting telescope images
are analyzed with commonly used methods, including a second-moment
analysis of the shower images in the cameras \cite{Hillas-1985}
and the reconstruction of shower directions and impact
parameters \cite{Kraw-2006}.

Extensive air showers induced by primary protons with energies
following a power law with a differential spectral index of -2.7 between 50 GeV and
10 TeV
are simulated with the CORSIKA code v.6.2 \cite{Heck-1998}.  As more
than 75\% of all cosmic rays are protons in this energy range and
arrays of IACTs are much less sensitive to air showers from heavier
nuclei (at the same energy),
only protons have been simulated (see as well section \ref{chap-pi0}).
The contribution from cosmic electrons has been neglected here.
The energy threshold after analysis cuts of the considered telescope array
is well above
100 GeV, where the cosmic electron flux is significantly below
the background due to hadronic cosmic rays.

The isotropic arrival directions of cosmic rays are simulated by
randomizing the shower directions in a cone with an opening angle of
$3.5^{\mathrm{o}}$ around the vertical pointing direction of the
telescopes. Shower cores are distributed randomly on a circular area
with a radius of 600 m around a point located roughly in the center of 
the array of telescopes.
The parameters for the
$\gamma$-ray simulations differ only in the differential spectral index (which
is -2.5, similar to the energy spectrum of the Crab Nebula \cite{Hillas-1998})
and the arrival
directions, which are assumed to be from a point source.  The shower
simulations use the U.S. standard atmosphere \cite{US-STANDARD} and atmospheric
extinction values estimated with Modtran 4 \cite{Kneizys-1996}, assuming 50 km visibility at
ground level for a wavelength of \mbox{550 nm}.

Two different combinations of low and high-energy interaction models are used:
QGSJet (version 01c) \cite{Kalmykov-1997} and FLUKA (version 2003.1)
\cite{Fasso-2005} with a transition
energy of 500 GeV, and Sibyll (version 2.1) \cite{Engel-1999} and
FLUKA (version 2003.1) with a transition energy of 
80 GeV\cite{Heck-2005b}.
Calculation of all electromagnetic interactions are performed with EGS4 \cite{Nelson-1985}
which is well tested and has small uncertainties, even up to 100 TeV.
Specifically, this is demonstrated by the good agreement
between simulated and real $\gamma$-ray images that is usually achieved for Cherenkov
telescope experiments (see e.g. ref. \cite{Maier-2005}).
Approximately $5\times 10^{8}$ proton events have each been simulated for the
QGSJet/FLUKA and Sibyll/FLUKA sets.

\begin{figure}
\centering\includegraphics[width=0.70\textwidth]{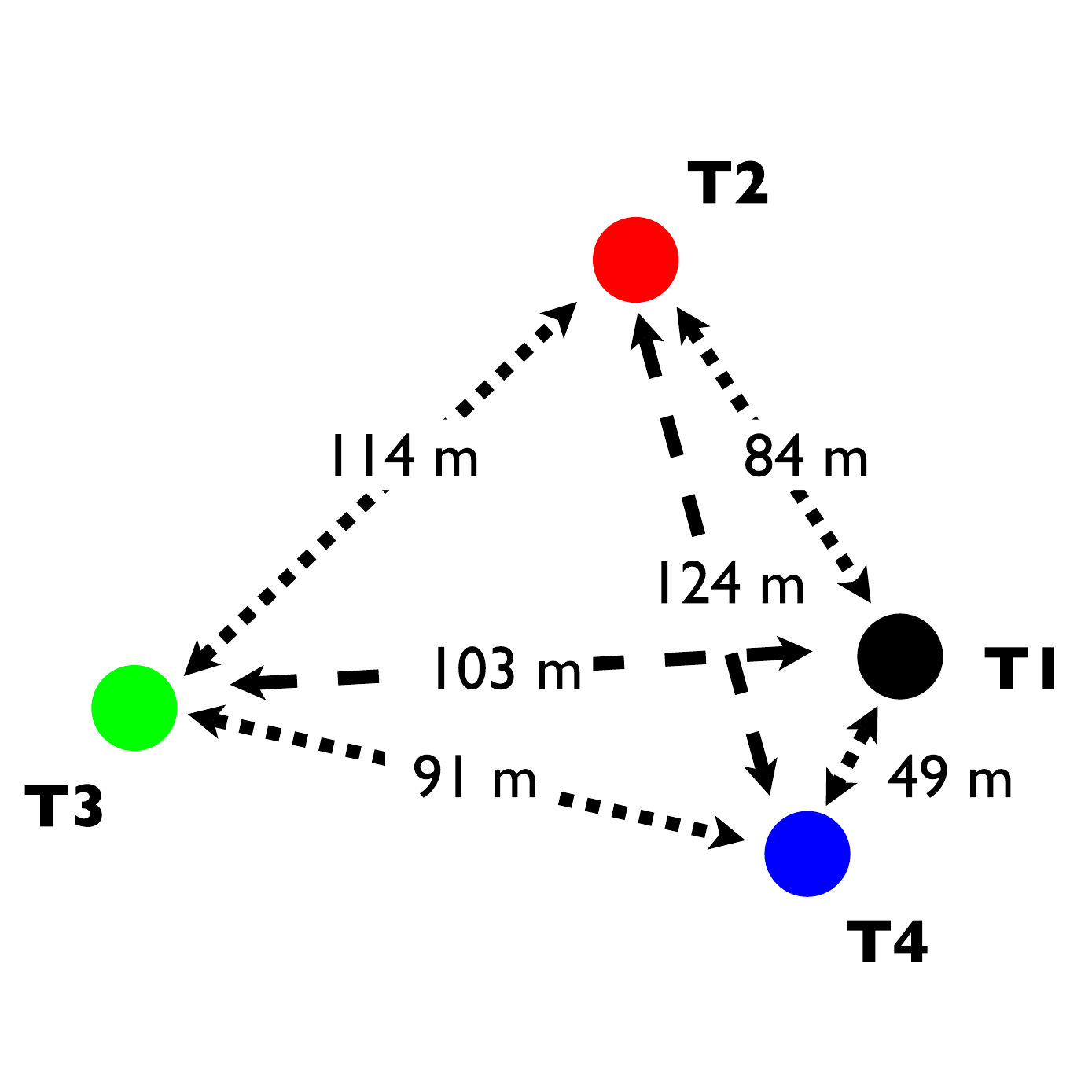}
\caption{\label{fig-telescopes}
Layout of the array of imaging atmospheric Cherenkov telescopes used in this paper
(similar to the temporary VERITAS layout).
}
\end{figure}

The array of IACTs consists of four telescopes arranged in a
quadrangle with different sides (see Figure
\ref{fig-telescopes}). The pointing direction of all telescopes is the
same, namely directly towards the source.  No convergent pointing towards
the shower maximum is used.  All telescopes are at the same altitude of
1270 m above sea level. The telescopes are Davies-Cotton
reflectors of 12 m diameter with a focal length of 12 m. Each
reflector comprises 350 hexagonal mirror facets of a total area of 110~m$^2$.
The cameras are equipped with $0.15^\circ$ diameter photomultiplier tubes (PMT) with
characteristics similar to the Photonis XP2970 model
in an hexagonal arrangement. Light cones are
simulated which yield a geometrical collection efficiency of 85\% for
Cherenkov photons hitting the focal plane. The local trigger system
consists of a simple multiplicity trigger of three adjacent PMTs with signals
above a threshold of $\approx4$ p.e.~in a time window of 5 ns. The array trigger requires
at least two telescopes with a local trigger in a time window of 100
ns.

The telescope simulation \cite{Duke} consists of two parts, the
propagation of Cherenkov photons through the optical system and the
response of the camera and electronics.  The signal in a PMT is
created by summing up single photo-electron pulses with appropriate
time and amplitude jitters.  Night sky background light, electronic
noise and all efficiencies, including mirror reflectivities,
geometrical, quantum, and collection efficiencies, and losses due to
signal transmission have been modeled.
The design of the telescopes and their arrangment
are similar to the temporary installation of the
VERITAS array \cite{Weekes-2002} at the Fred Lawrence Whipple
Observatory in Southern Arizona, USA.
The presented work was prepared during the construction phase of VERITAS and 
several changes in layout and design of the actual system made 
it impossible to compare in detail the simulations presented here with array 
data. The simulation chain has been extensively tested on single 
telescope data and reproduces the characteristics of the VERITAS-1 
telescope \cite{Maier-2005}.

The event reconstruction procedure consists of image cleaning,
second-moment image analysis \cite{Hillas-1985} for each camera, and
reconstruction of shower direction and impact parameter on the ground,
using all available images.  Images are cleaned by removing pixels
with signals of less than five and three times the
noise variation for 
pixels near the centre and at the perimeter of the image,
respectively. 
In the second moment analysis, the axes and widths of the image 
are determined. Images of at least five pixels
are required for the array reconstruction. The intersection of the
extrapolated long axes of the different images defines the shower
direction \cite{Hofmann-1999}.

Thus, events first have to fulfill a set of {\em event quality} conditions: 
at least 5 pixels per image, the moment analysis must be 
successful, and reconstruction of the shower direction and 
impact point must be possible. 

\begin{figure}[p]
\centering\includegraphics[width=0.70\textwidth]{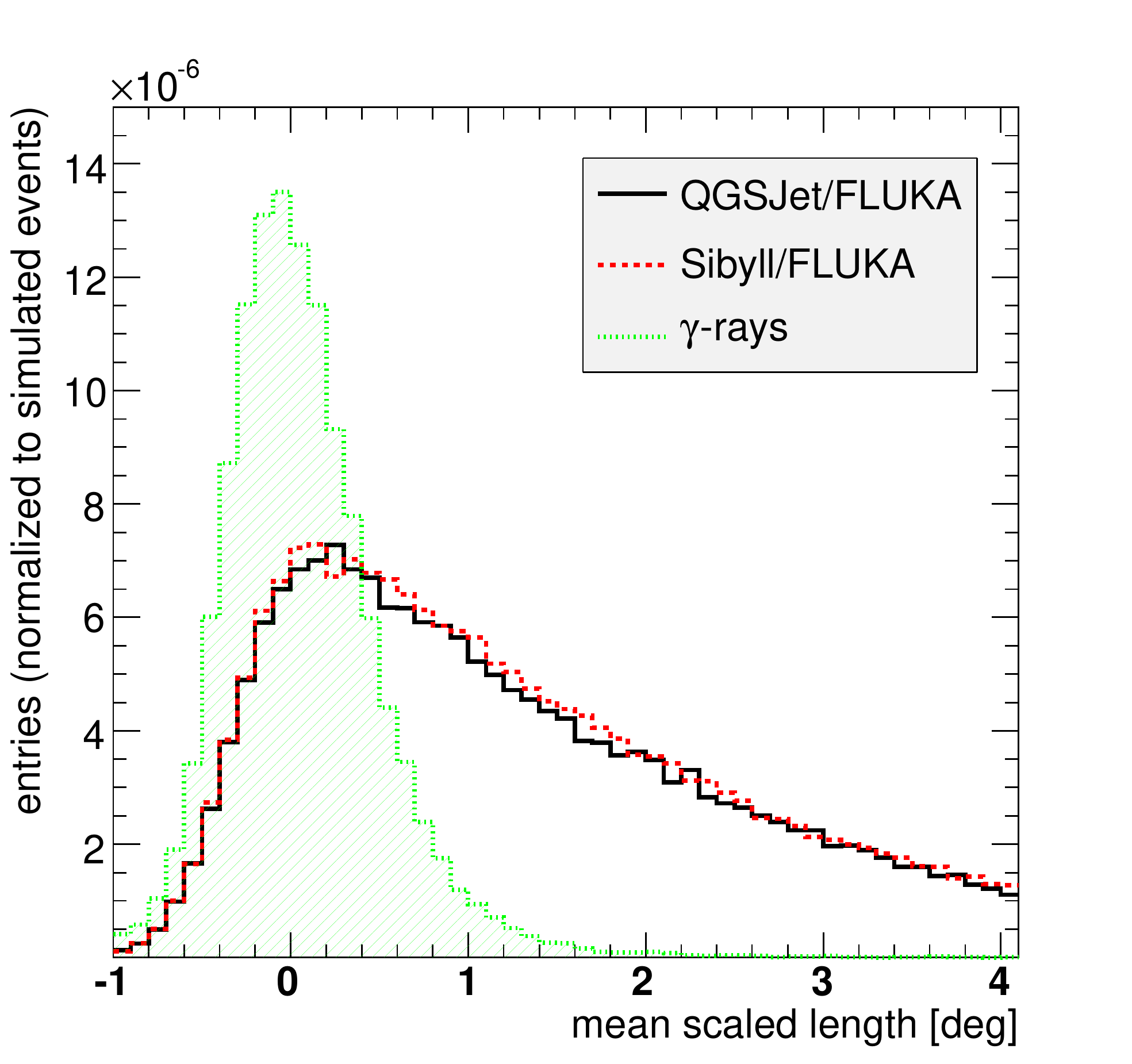}
\caption{\label{fig-mscw}
Mean scaled width distribution for events passed the trigger and reconstruction cuts.
Primary $\gamma$-rays and protons are shown
(QGSJet/FLUKA and Sibyll/FLUKA simulations).
}
\end{figure}

\begin{figure}[p]
\center\includegraphics[width=0.70\textwidth]{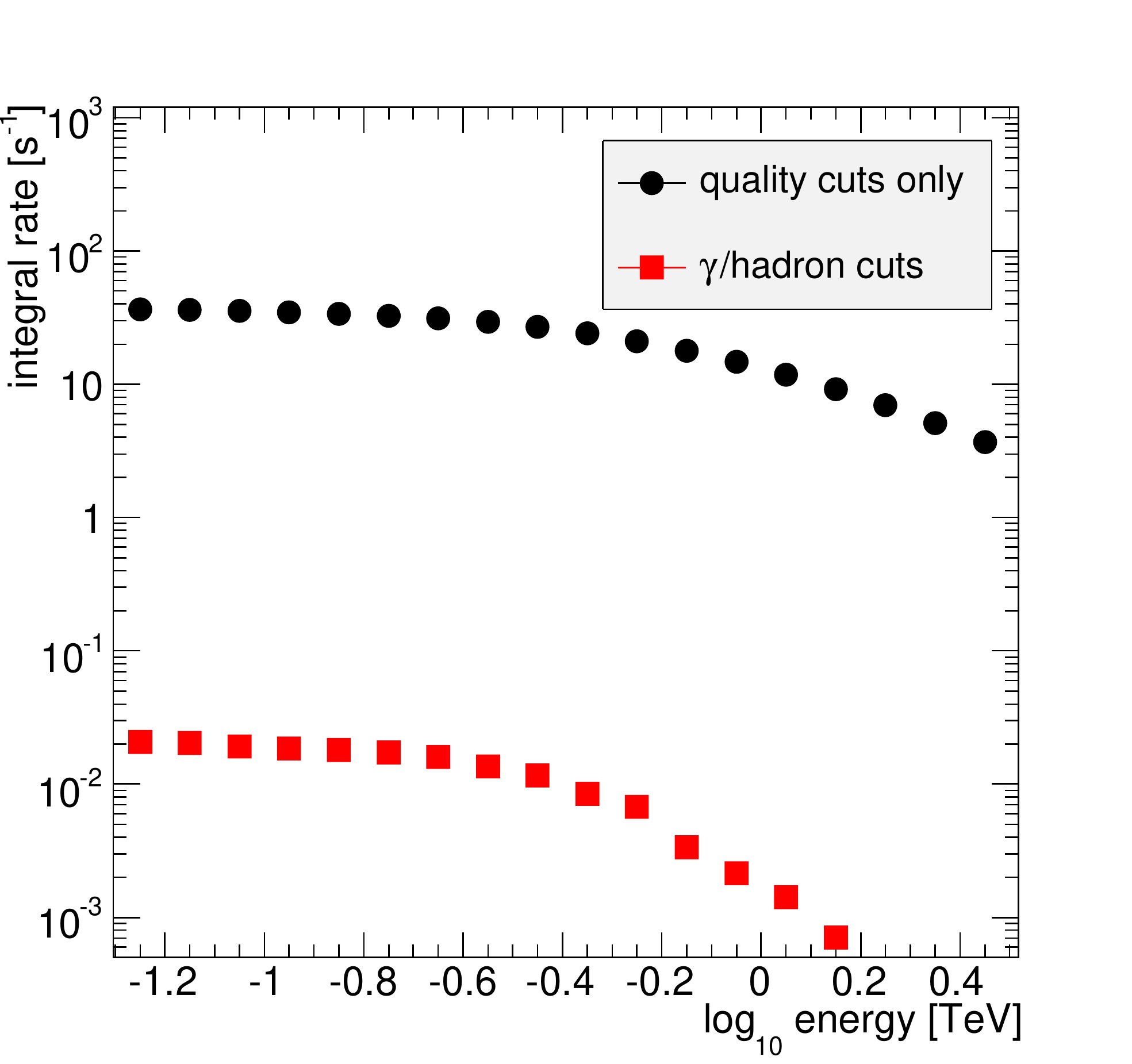}
\caption{\label{fig-trigger}
Integral rates of primary protons after quality cuts, and after $\gamma$-hadron separation cuts
(QGSJet/FLUKA simulations).
}
\end{figure}

The background rejection method consists of shape and direction cuts.
Showers induced by cosmic rays usually have much broader and very
irregular images in the camera, while showers induced by $\gamma$-rays
have the shape of slim, elongated ellipses.
Look-up tables containing the expected width and length for a given image size and
distance of the telescope from the shower impact position ($width_{expected} (R_i, size_i)$)
are filled from simulated $\gamma$-ray showers.
These tables are used to
calculate the mean deviation in width and length of images in each
camera from their expected values \cite{Aharonian-1997}.
These mean values are generally called {\em mean scaled width}
(MSCW) and {\em mean scaled length} (MSCL).  MSCW is defined according
to
\begin{equation}
\mathrm{MSCW} = \frac{1}{N_{tel}} \times \left[ \sum_i^{N_{tel}} 
\frac{width_i - width_{expected} (R_i, size_i)}{\sigma_{width_{expected}} (R_i, size_i)} \right]
\end{equation}
where $N_{tel}$ is the number of telescopes with a good image, $R_i$
is the distance of telescope $i$ from the shower impact position,
$size_i$ is the total number of photoelectrons recorded in telescope
$i$. An analogous formula is used to calculate mean scaled length.
Figure \ref{fig-mscw} shows the significant difference between the MSCW
distributions for $\gamma$-rays and protons.



The last important cut is on the direction of the incoming shower.
Only events coming from the source direction are accepted as $\gamma$-rays.
This is expressed through the variable $\Theta^2$, which describes
the squared angular difference between source and reconstructed shower direction.
Obviously
this cut suppresses a significant part of the isotropic background,
while retaining the source events. 
The better the angular resolution of the telescope array,
the more restrictive this cut can be chosen. 

The following $\gamma$-hadron selection cuts have been used: candidate
events are accepted as $\gamma$-rays if MSCW $< 0.3$, MSCL $< 0.45$, and
$\Theta^2 < 0.015$ deg$^2$.
(To increase the number of $\gamma$-like proton events a much wider
direction cut for proton events
of $\Theta^2 < 1$ deg$^2$ is applied in the following.
All results are then scaled to the opening angle of $\Theta^2 < 0.015$ deg$^2$.).

Table \ref{tab-events} shows that only about one in seven million
simulated proton showers passes the trigger and reconstruction requirement and
the $\gamma$-hadron separation cuts
(this value depends on the size of the scatter area).
With the described $\gamma$-hadron selection cuts
a hadron suppression factor of about 2000 relative
to the number of reconstructed proton events is achieved.
In contrast, one out of 200 simulated $\gamma$-rays passes these cuts,
about 60\% of all reconstructed $\gamma$-rays are lost in the process.
The term
{\em $\gamma$-like event} is used in the following for an event which
passed all cuts described above.


Figure \ref{fig-trigger} shows the integral rates for protons
for quality cuts only (at least 5 pixels per image, successful image
parameterisation, and reconstruction of shower direction and impact point) and
after $\gamma$/hadron separation cuts have been applied.

\section{A closer look at $\gamma$-like proton showers}
\label{chap-background}

The proton events which pass all $\gamma$-hadron
separation cuts, described in the previous section, end inevitably in
the $\gamma$-ray sample. It is known that the decay of high-energy
neutral pions into two $\gamma$-rays in an early stage of the shower
development is often the reason for proton showers to look like
$\gamma$-ray showers (see e.g.~\cite{Schatz-1996}).

\begin{table}[p]
\begin{center}
\begin{tabular}{|c|c|c|c|c|c|c|c|}
\hline
          &  & number of & fraction of & fraction of &  & & \\
          &  & simulated & reconstructed& $\gamma$-like & $N_{tel} = 2$ & $N_{tel} = 3$ & $N_{tel} = 4$ \\
          &  & events & events & events &  & & \\
\hline
QGSJet/FLUKA & All events & $4.6 \cdot 10^{8}$ & $ 3.3 \cdot 10^{-4}$ & $1.5 \cdot 10^{-7}$ & 83\% & 13\% & 4\% \\
(proton & $F_{\mu} < 0.5$ & - & - & 70\% & 65\% & 99\% & 100\% \\
simulations) & $F_{\mu} > 0.5$ & - & - & 30\% & 35\% & 1\% & 0\% \\
\hline
Sibyll/FLUKA & All events & $5 \cdot 10^{8}$ & $ 3.4 \cdot 10^{-4}$ & $1.5 \cdot 10^{-7}$ & 83\% & 13\% & 4\% \\
(proton & $F_{\mu} < 0.5$ & - & - & 72\% & 67\% & 99.5\% & 100\% \\
simulations) & $F_{\mu} > 0.5$ & - & - & 28\% & 33\% & 0.5 \% & 0\% \\
\hline
$\gamma$-rays & All events & $5\cdot 10^{6}$ & $1.2 \cdot 10^{-2}$ & $5\cdot 10^{-3}$ & 6.9\% & 26.4\% & 66.7\% \\
\hline
\end{tabular}
\end{center}
\caption{\label{tab-events} Overview of number of simulated and
  reconstructed events, number of selected events, and
  fraction of events with telescope multiplicity 2, 3, and 4.
  Note that the rows for $\pi^{0}$-like ($F_{\mu}<0.5$) and muon-like
  events ($F_{\mu}>0.5$) quantify the number of 2, 3, or 4-telescope
  events as fraction of all events with the same number of telescopes.
  Selected events are events with MSCW $< 0.35$, MSCL $<0.45$, and 
  $\Theta^{2}< 0.015$ deg$^2$.
}
\end{table}

\begin{figure}[p]
\includegraphics[width=0.49\textwidth]{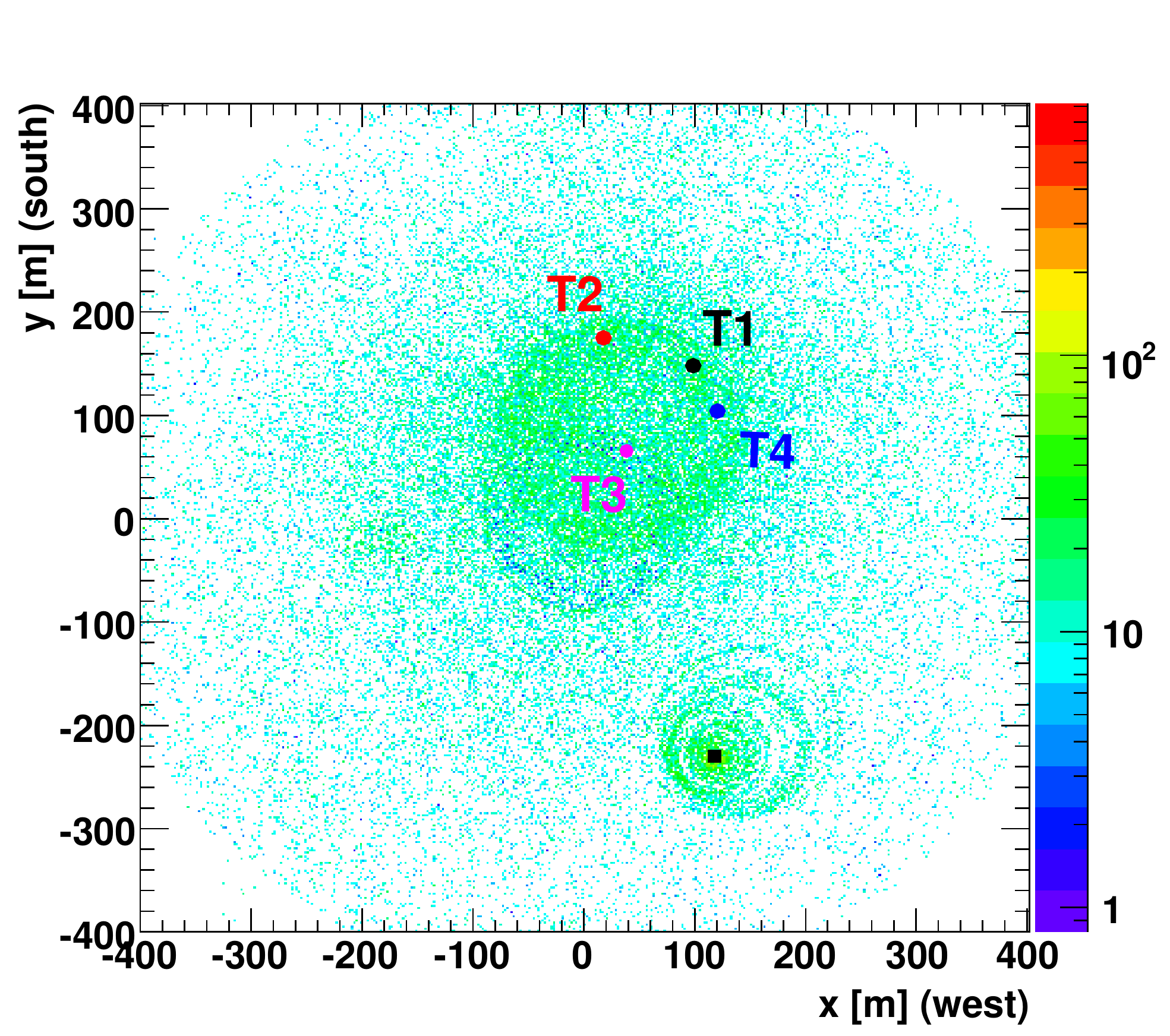}
\includegraphics[width=0.49\textwidth]{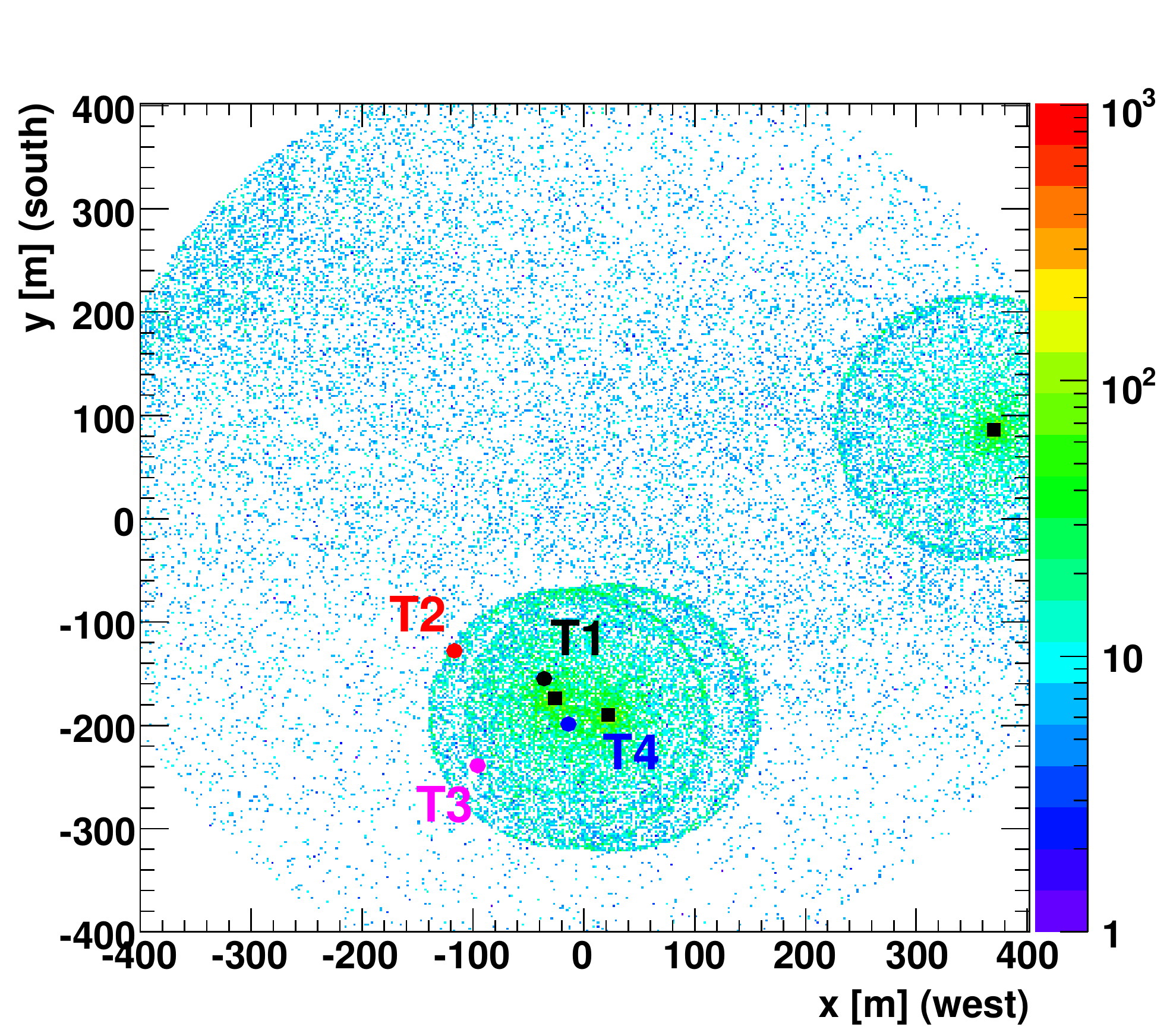}
\caption{\label{fig-xy}
Cherenkov photon distributions at ground for two typical $\gamma$-like
events. The color scale indicates the number of Cherenkov photons per
bin.
left: telescopes triggered by pionic subshower.  right:
telescopes triggered by muon rings.  The position of the four
telescopes are indicated by small circles (T1: black, T2: red, T3:
green, T4: blue).  High-energy muons ($E_{\mu}>$ 5 GeV) are drawn as
small black squares.
Only Cherenkov photons with distances smaller than 450 m to the
position of the shower core are shown. (QGSJet/FLUKA simulations)
}
\end{figure}

However, in our event sample we find a second class of $\gamma$-like 
events.  These are events with high-energy muons ($>$ few GeV)
near the array center.
Cherenkov light emitted from these muons is usually confined to a much
smaller area than Cherenkov light from air showers, but high local
photon densities can trigger telescopes and display $\gamma$-like
images.  Due to the small lightpool of an energetic muon, multiple
telescope images occur only if telescopes are close to one another.  The array
layout considered here, with distances between individual telescopes
as small as 49 m, causes the relatively large sensitivity to such
muonic events.  Generally, distances between telescopes are much
larger, in the range of 80 m (e.g.~the future VERITAS array or
MAGIC-II) to 120 m (e.g.~H.E.S.S.), and consequently the muon
sensitivity is reduced for those arrangements.


\subsection{Muonic $\gamma$-like proton showers}
\label{chap-muons}

The total number of Cherenkov photons emitted by a muon and their spatial distribution
depends mainly on the muon energy
(the number of Cherenkov photons $N_C$ produced per path length $s$ is
$\mathrm{d}N_C/\mathrm{d}s \propto \int \sin^2 \Theta_C/\lambda^{2} \mathrm{d}\lambda$.
$\Theta_C$ describes the Cherenkov emission angle $\cos \Theta_C = 1/(\beta n)$,
 $\beta$ the muon velocity, $n$ the refractive index, and 
$\lambda$ the wavelength of the Cherenkov photons).
Most of the light seen by the telescopes is produced within a few kilometers
of ground level.

In proton showers of 50 GeV to 10 TeV, typical muon energies 
at ground level
are between 1 and 20 GeV, but the muon energy distribution
extends beyond several hundred GeV.  These muons originate mainly
from low-energy protons which produce only
one or a few secondaries in a peripheral collision, 
which then produce one or a few muons,
carrying most of the primary energy.  The few other shower particles
are absorbed at large heights, and almost no particles or Cherenkov
photons reach the ground.  Figure \ref{fig-xy} (right) shows a typical
muonic event which passed all $\gamma$-hadron separation cuts.
Several circular Cherenkov light pools of about 120 m radius can be
seen, corresponding to muon energies of roughly 50 GeV to 80 GeV.
This muonic event triggered telescopes 1 and 4, which have the smallest
distance to each other (49 m).

The role of Cherenkov light from muons was studied with a special
set of simulations. All $\gamma$-like showers were simulated a
second time, but with Cherenkov emission from muons now switched off. By comparing
the two simulation sets event-by-event, the fraction of Cherenkov photons hitting
a telescope which were produced by muons (=$F_{\mu}$) could be determined.

%
%
\begin{figure}[p]
\centering\includegraphics[width=0.70\textwidth]{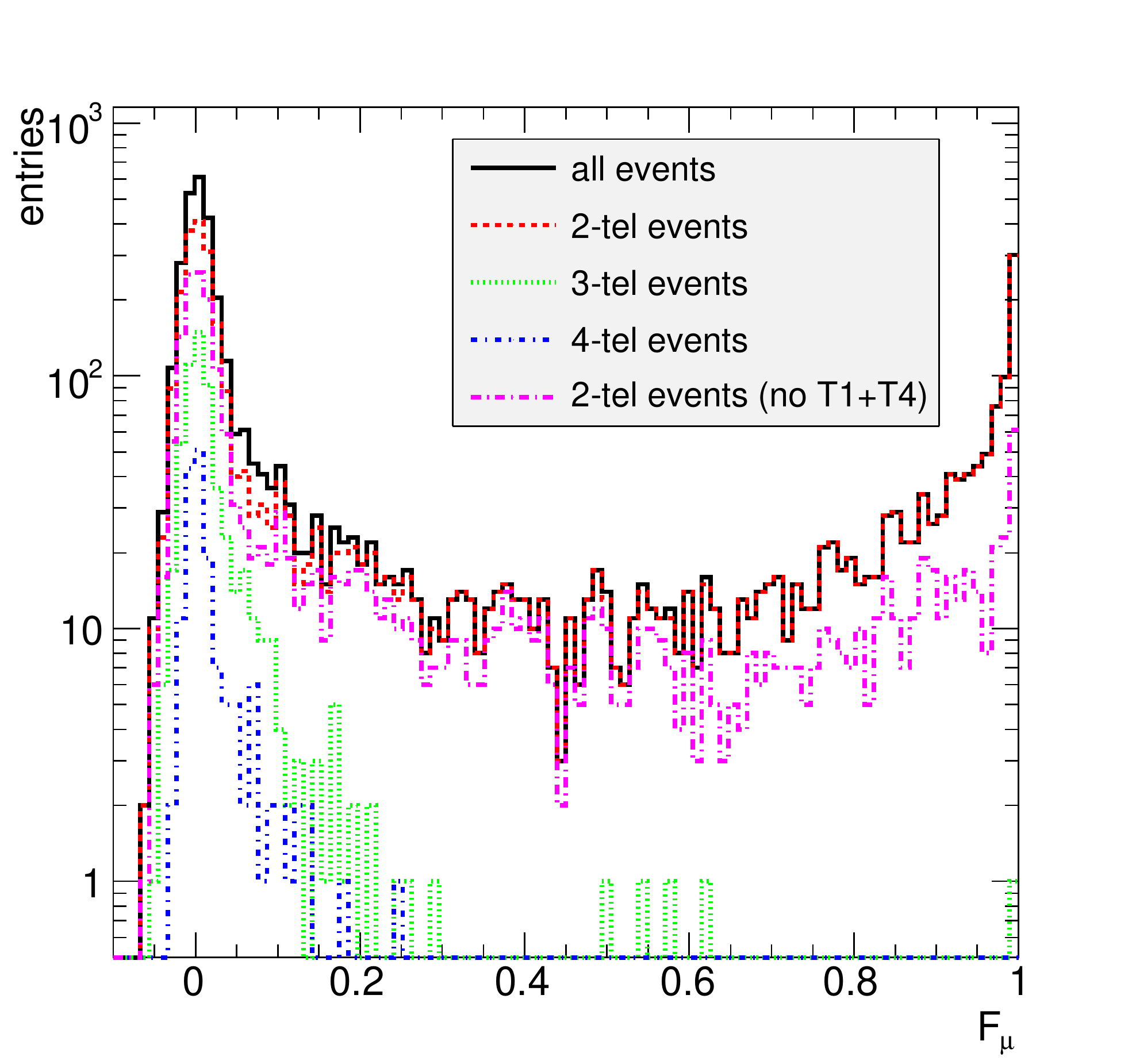}
\caption{\label{fig-sc-tel}
Distribution of the fraction of measured Cherenkov light emitted
by muons ($F_{\mu}$) in $\gamma$-like events for different telescope
multiplicities
(QGSJet/FLUKA simulations).
}
\end{figure}

%
\begin{figure}[p]
\centering\includegraphics[width=0.70\textwidth]{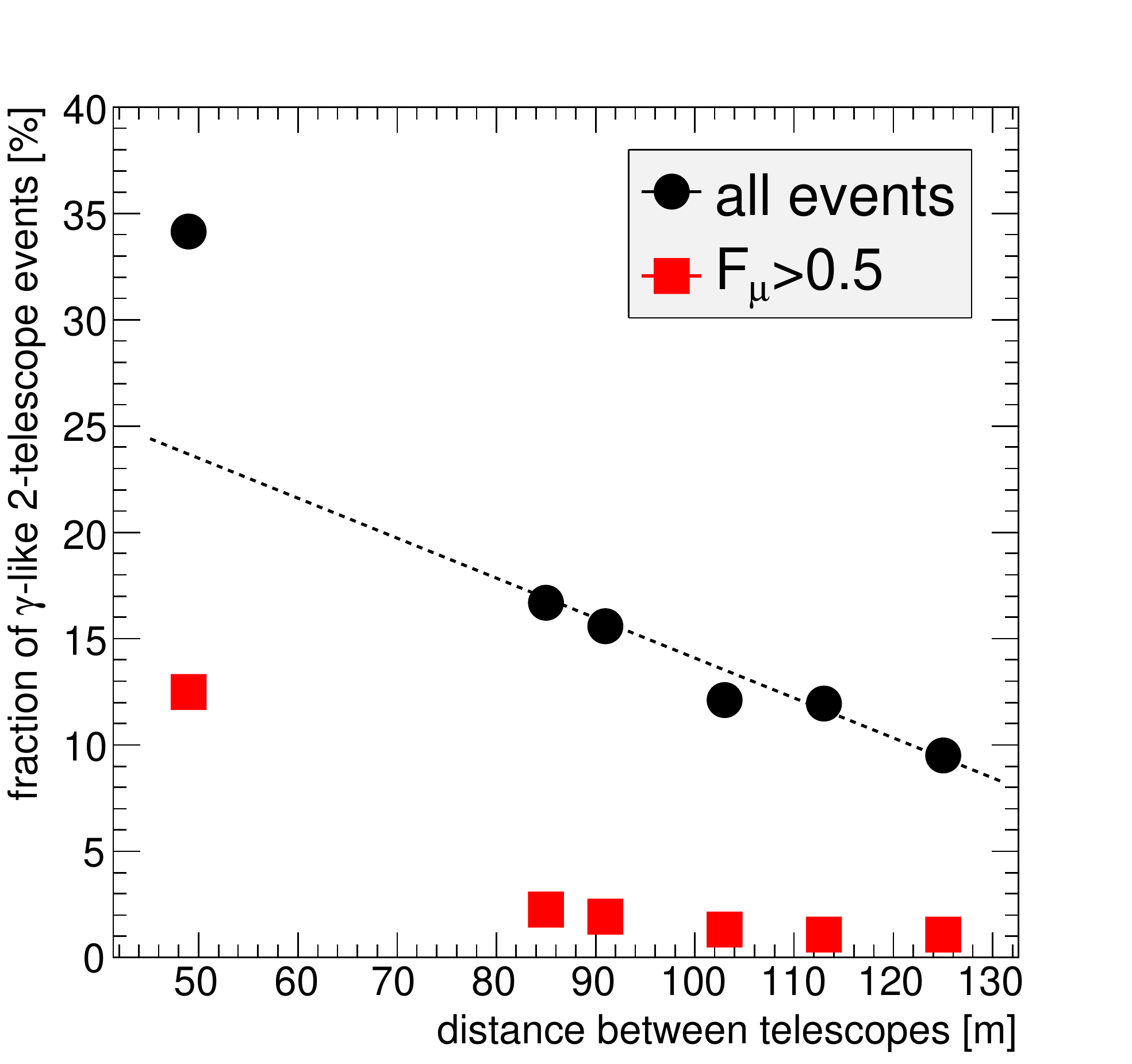}
\caption{\label{fig-sc-distance}
Fraction of $\gamma$-like events with a two-telescope array
trigger vs.~distance between each telescope pair. The dotted line is
drawn to guide the eye
(QGSJet/FLUKA simulations).
}
\end{figure}

$F_{\mu}$ is plotted for all $\gamma$-like proton events in Figure
\ref{fig-sc-tel}. The fraction of events with $F_{\mu} < 0.5$ and
$F_{\mu} > 0.5$ are listed in Table \ref{tab-events}. $F_{\mu} > 0.5$
means that more than 50\% of all Cherenkov photons recorded by the
telescopes is emitted by muons. It is obvious that
there are two classes of events and that muonic showers, i.e.~showers
with $F_{\mu} > 0.5$ are only able to trigger two telescopes. In
about 30\% of all $\gamma$-like proton showers, Cherenkov photons
from muons are the dominant part. Almost all of them are two-telescope
events. Events with $F_{\mu} > 0.5$ are called {\em muonic events} in
the following.

The number of $\gamma$-like two-telescope events increases sharply
with shrinking telescope distances due to the contribution of muons
(Figure \ref{fig-sc-distance}). 
The rate increase can be understood as the increase in the
allowed region of impact points for muons with a given energy, and
hence radius of Cherenkov light pool. The area of this
allowed region is small or zero for small muon energies or large
telescope distances.  Figures \ref{fig-sc-tel} and \ref{fig-sc-distance}
show that more than half of the
$\gamma$-like muonic events are T1+T4 events, the pair with the
smallest telescope distance (49 m).

There are several ways to eliminate the majority of these events.  The
most obvious ones are to modify the trigger condition 
to use only 2-telescope triggers of 
pairs with larger distances (e.g. pairings other than Telescope T1+T4) 
or to require $>2$ triggered telescopes. This suppresses
muonic events at the trigger level (rejection of 2-telescope triggers on T1+T4 reduces the
number of muonic $\gamma$-like events by about 40\%), but also reduces the sensitivity to
primary $\gamma$-rays at low energies.  Another approach uses the
difference in the emission height of the Cherenkov photons 
\cite{Aharonian-1997} of muonic $\gamma$-like events and $\gamma$-ray
showers.  The production height of Cherenkov photons from muons 
(in muonic 2-telescope events) 
is typically below 2-3 km above ground.
In contrast, Cherenkov light from air showers is emitted around
the shower maximum at typically 8-10 km height.

\begin{figure}[p]
\centering\includegraphics[width=0.60\textwidth]{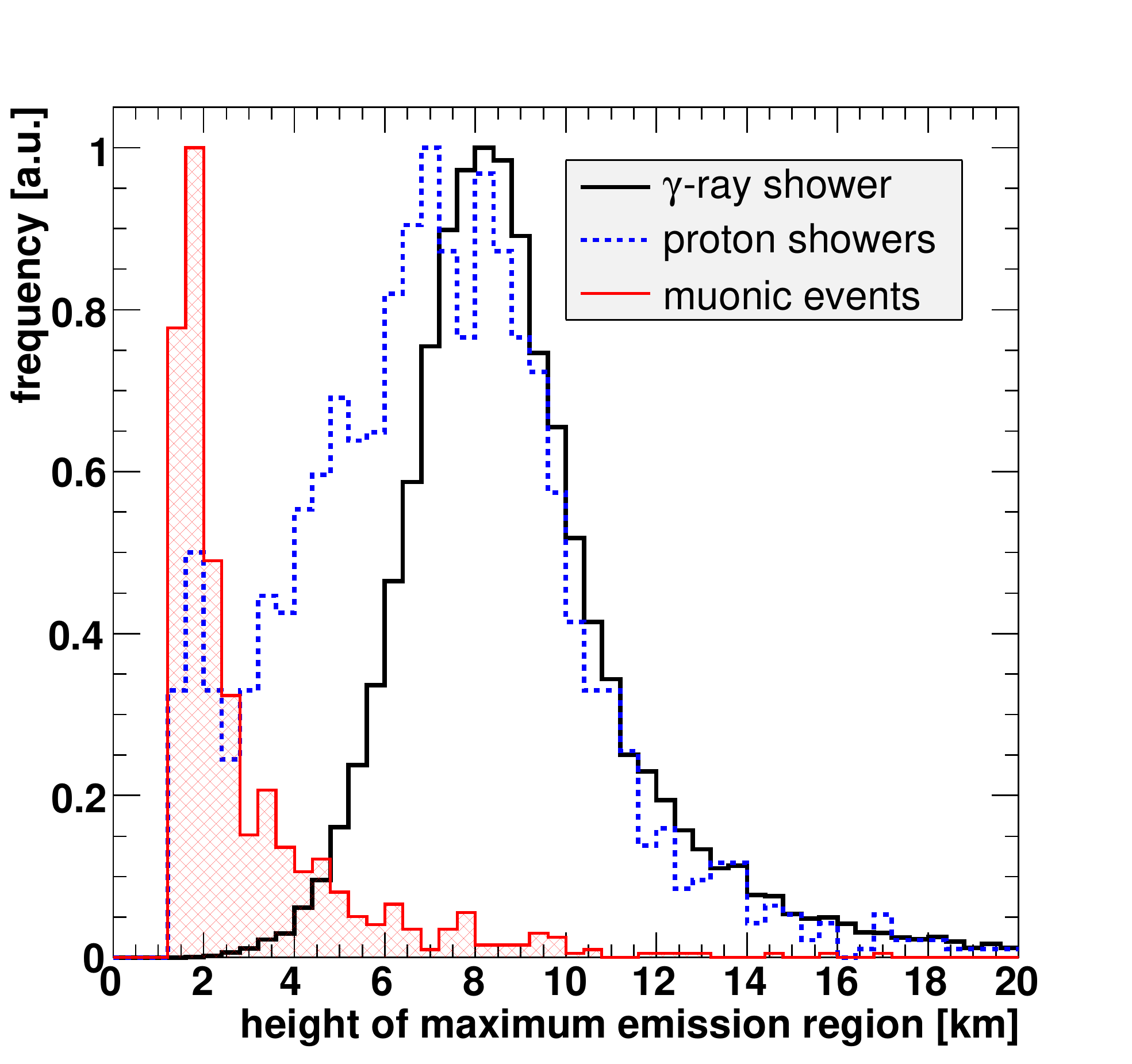}
\caption{\label{fig-muons}
Height of the region of maximum Cherenkov photon emission, calculated from the position of the
image centroids in Telescopes 1 and 4 
(QGSJet/FLUKA simulations).
}
\end{figure}

The distance $c$ between the image centroids in the cameras of two
telescopes pointing towards the source (parallel pointing mode) is
related to the distance $D$ of the telescopes to each other and the
height of the Cherenkov emission maximum $h$ by $c = D / h$.  Figure \ref{fig-muons}
shows the distributions of
$h$, estimated with this simple relationship for $\gamma$-rays and
muonic $\gamma$-like events. There is a very clear separation between
the two distributions, and with a cut at \mbox{$h = 4$ km} about
80\% of all muonic $\gamma$-like events can be suppressed, while
less then 2\% of the $\gamma$-ray events are lost.


\subsection{Pion-induced $\gamma$-like proton showers}
\label{chap-pi0}

%
%
%

\begin{figure}[p]
\centering\includegraphics[width=0.70\textwidth]{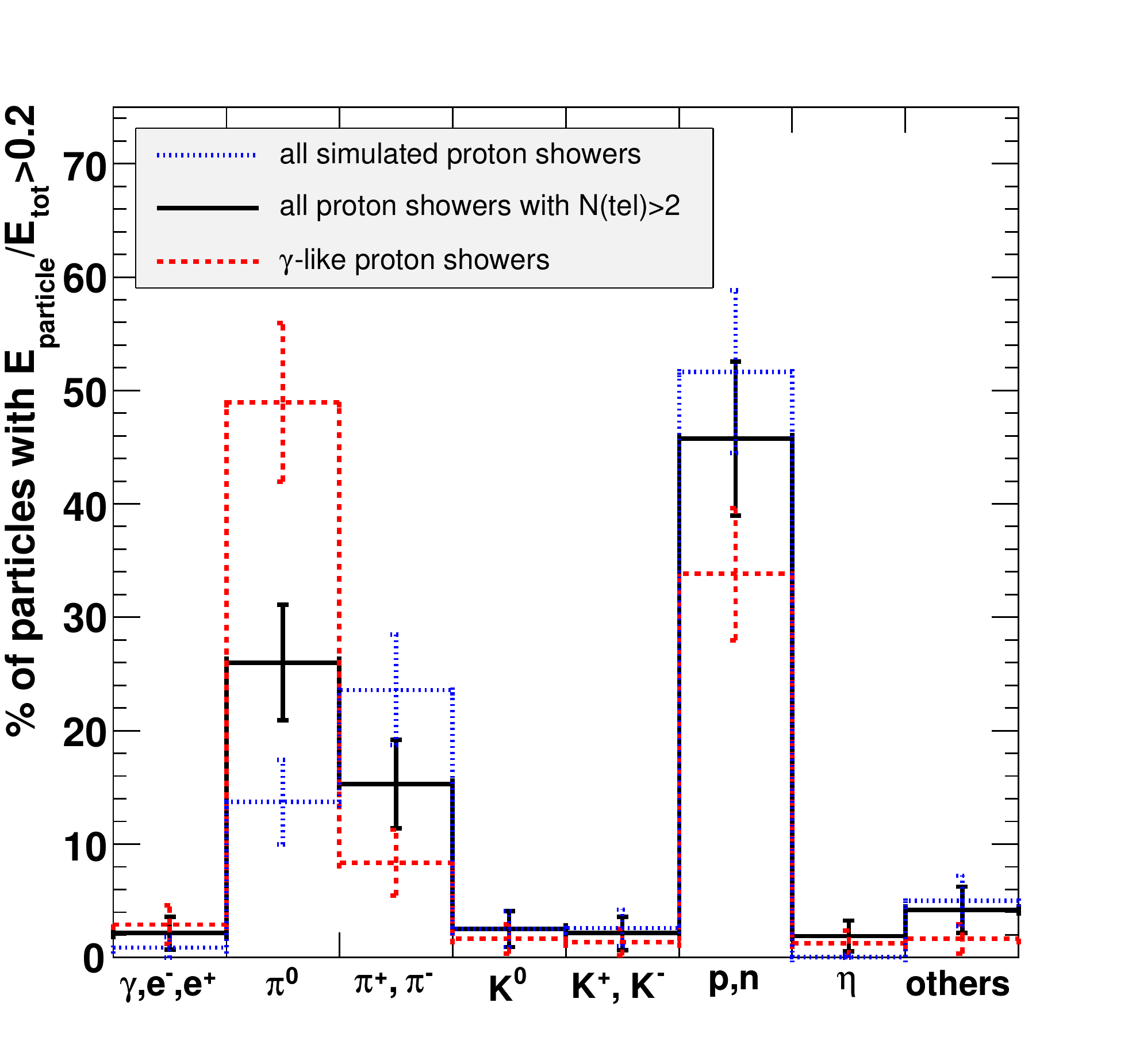}
\caption{\label{fig-fid-sec}
Distribution of secondary particles from the first interaction in proton showers.
Only particles with energies larger than 20\% of the primary energy are counted
(QGSJet/FLUKA simulations).
}
\end{figure}

A proton shower is approximately a superposition 
of many electromagnetic subshowers initiated by the
decay of neutral pions, and of muons from the decay of charged pions.
The different subshowers produce an irregular Cherenkov photon
distribution at the ground, and thus the images of proton showers in the camera of
an IACT are usually patchy and broad. This makes most of them easily
distinguishable from $\gamma$-ray showers. For a $\gamma$-like
image, a proton shower must either be dominated by one subshower or 
only one of the subshowers is seen by the telescopes (see example in Figure
\ref{fig-xy} left). Therefore, in the following,
particle production in the early shower development is investigated, 
and especially
those secondary particles which carry a significant part of the
primary energy.  
Events with a 3-fold array trigger are selected
to remove all muon-induced events from the simulated event sample.

Figure \ref{fig-fid-sec} shows the distribution of secondary
particles 
in the first interaction in proton showers.  Only particles 
with an energy of at least 20\% of the primary energy are counted.
While the distribution for all simulated events shows the expected ratio of
charged to neutral pions of 2 to 1, this ratio approximately reverses for 
particles in events with a 3-fold array trigger.  
Well above-average Cherenkov light emission 
is needed to trigger three or more telescopes.  
As the large number of $\pi^{0}$'s indicates, 
the light originates in electromagnetic subshowers
initiated by $\pi^0$ decay. About 50\% of all secondaries in the first
interactions of $\gamma$-like proton showers are neutral pions.
The second neutral particle with predominately electromagnetic decay
modes is the $\eta$-meson.  While it constitutes only a very small
fraction of secondaries in normal proton showers, a few percent of
$\gamma$-like events with a 3-fold array trigger contain high energy $\eta$'s. 


%
\begin{figure}[p]
\centering\includegraphics[width=0.70\textwidth]{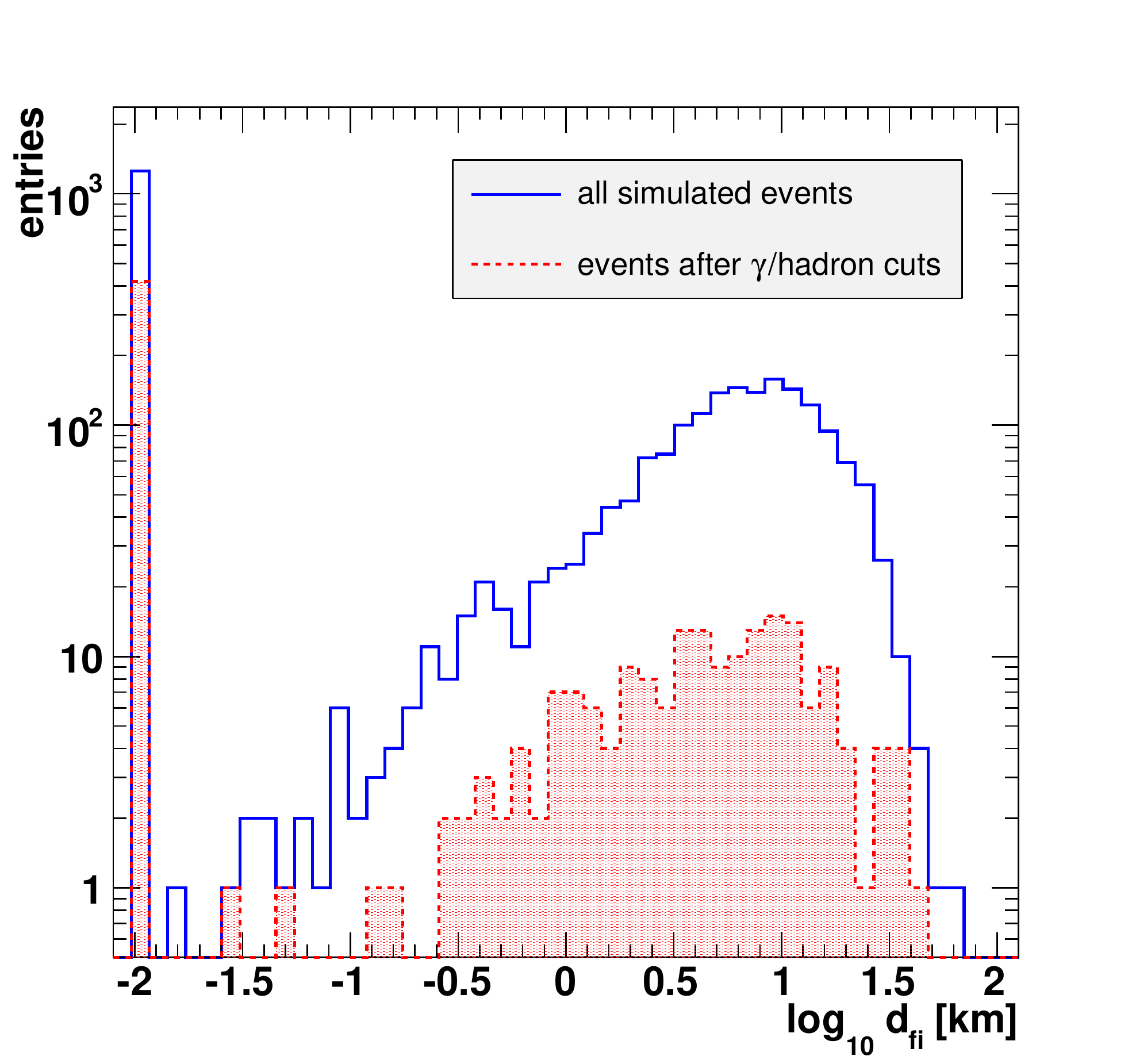}
\caption{\label{fig-pi0-2a}
Distribution of distance $d_{fi}$ between height of first interaction and $\pi^{0}$
production during shower development for events with $E_{\Sigma}(\pi^{0})/E_{tot}>0.3$ .
Events with $d_{fi} < 10$ m are entered in the bin at $\log_{10} d_{fi}=-2$
(Sibyll/FLUKA simulations).
}
\end{figure}


Simulations of $\gamma$-induced showers show that an energy of about
80 GeV or more is needed to trigger the array and pass all
quality and $\gamma$-hadron separation cuts. Similar energies are 
required too, in the dominating subshower for $\gamma$-like proton
showers. To investigate this, 
in the simulations
the energy and production height of all
$\pi^{0}$'s and $\eta$'s with energies above 5 GeV have been recorded. 
Two variables are examined further, the total energy
sum in $\pi^{0}$'s or $\eta$'s (E$_{\Sigma}(\pi{^0},\eta)$) and the
fraction of the primary energy carried away by these particles
((E$_{\Sigma}(\pi{^0},\eta)/$E$_{tot}$).  Figure \ref{fig-pi0-2a} 
shows the distribution of the distance between the height of first
interaction and the $\pi^{0}$ production during shower development for events with
E$_{\Sigma}(\pi{^0})/$E$_{tot} > 0.3$.  The figure suggests three
categories of events: (i) events with
E$_{\Sigma}(\pi{^0},\eta)/$E$_{tot} > 0.3$ in or close to first
interaction, (ii) E$_{\Sigma}(\pi{^0},\eta)/$E$_{tot} > 0.3$ later
during the shower development, and (iii)
those events which do not appear at all in the figure (E$_{\Sigma}(\pi{^0},\eta)/$E$_{tot} < 0.3$ for the whole shower).
In
this example, about 50\% of the relevant particle production occurs in
or shortly after the first interaction, about 25\% of the events pass
this threshold about 5 to 10 km after the first interaction, and 25\%
of the events never accumulate 30\% of their energy in the electromagnetic channel.

\begin{figure}[p]
\includegraphics[width=0.49\textwidth]{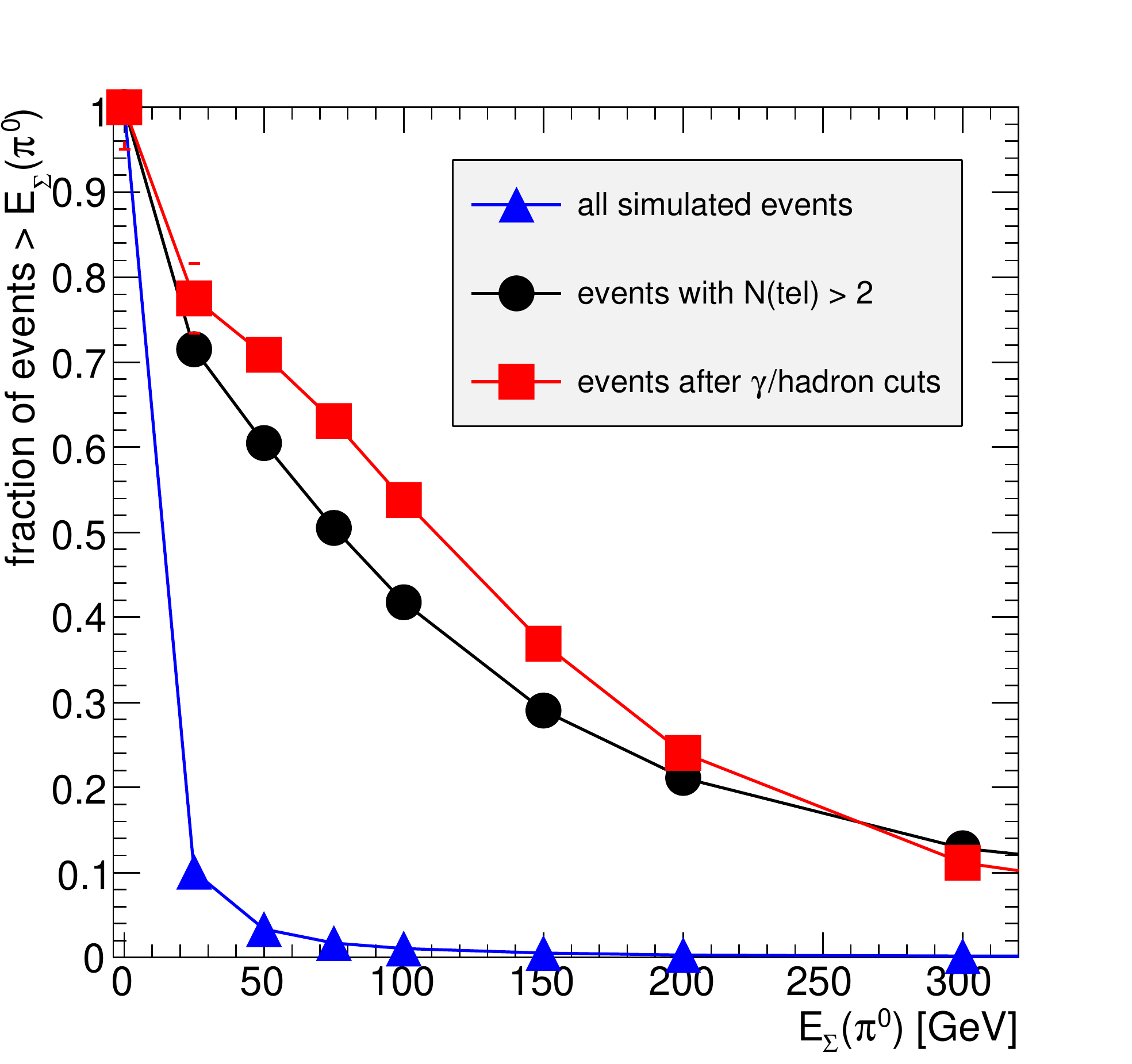}
\includegraphics[width=0.49\textwidth]{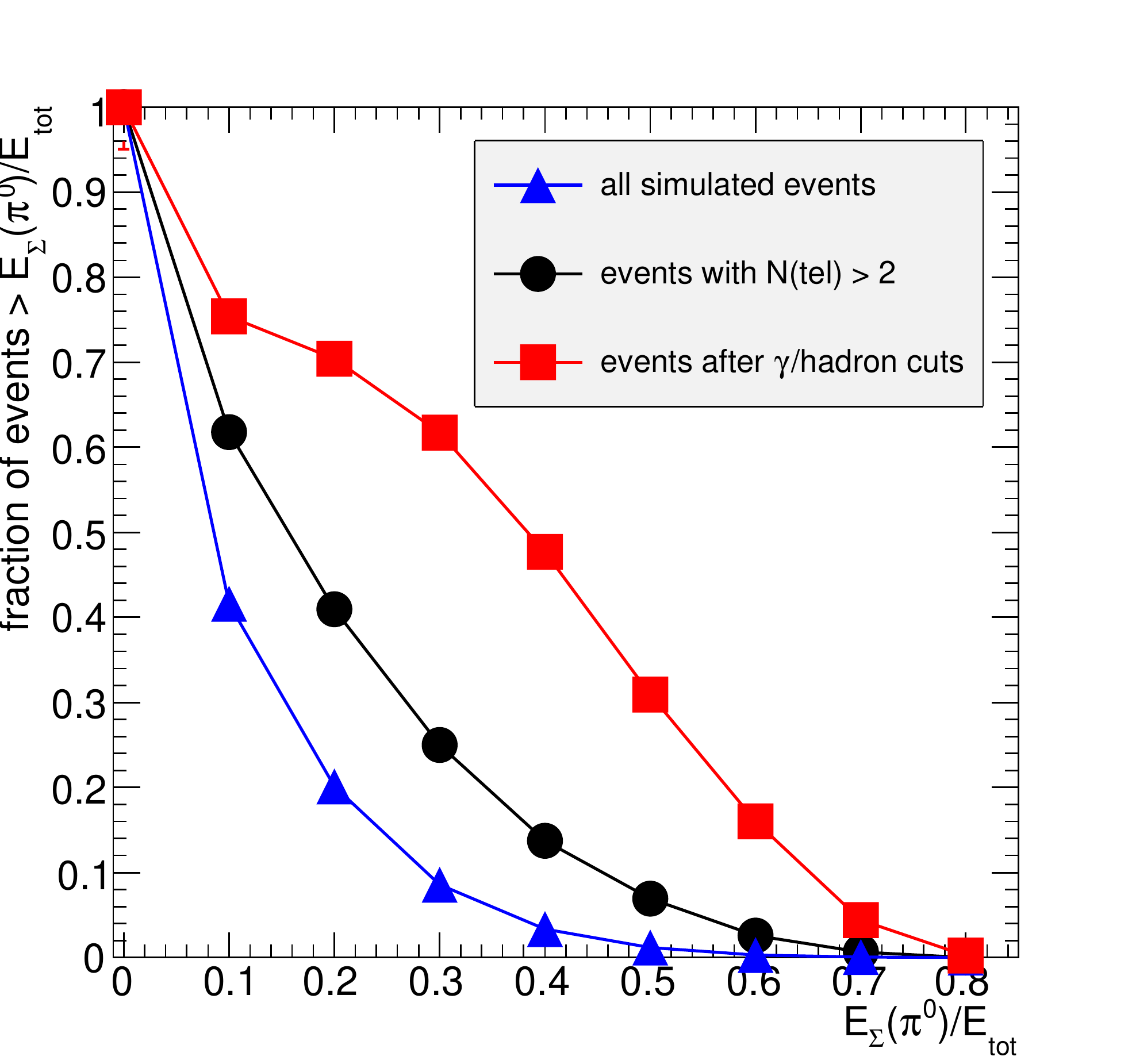}
\caption{\label{fig-pi0-3}
Fraction of events with $E_{\Sigma}(\pi^{0})$ (left) and $E_{\Sigma}(\pi^{0})/E_{tot}$ (right)
larger than the value on the abscissa in, or close to, the first interaction 
($d_{fi}< 10$ m).
All values for Sibyll/FLUKA simulations.}
\end{figure}

The importance of the production of high-energy $\pi^{0}$'s is
highlighted in Figure \ref{fig-pi0-3} (left and right). The fraction
of events with a $\pi^{0}$ energy sum above 80 GeV in all simulated
proton showers is very small, while the majority of events with a 3-fold
array trigger exceed this threshold. $\gamma$-like events are even
more likely to contain high-energy $\pi^{0}$'s.  The dominance of
electromagnetic subshowers in $\gamma$-like proton events can be seen
in Figure \ref{fig-pi0-3} (right).
These events are 4-10 times more likely to have an electromagnetic energy share of
40\% or more of the primary energy.

Both findings suggest that, firstly, the electromagnetic part in the
proton initiated shower has to be energetic enough to trigger the
array and, secondly, this part must carry a significant part of the
primary energy to prevent the occurrence of other large subshowers
that would disturb the $\gamma$-like appearance of the 
Cherenkov image in the cameras.

\begin{figure}[p]
\centering\includegraphics[width=0.60\textwidth]{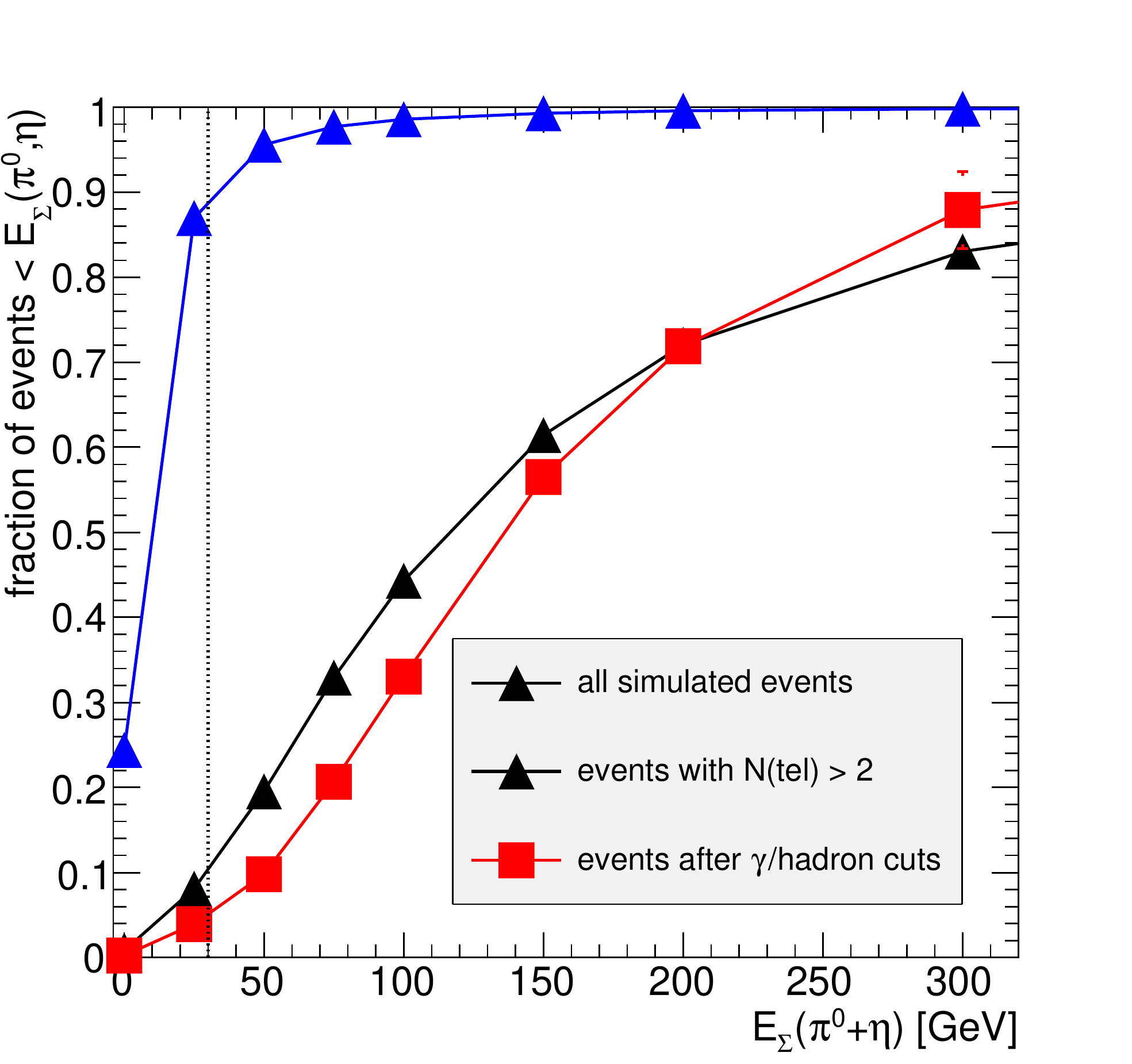}
\caption{\label{fig-pi0-2b}
Fraction of events with E$_{\Sigma}(\pi{^0},\eta)$ (sum of energy of all $\pi^{0}$'s and $\eta$'s)
smaller than the value on the abscissa during the whole shower development.
Events with $N_{tel}\geq$ 3 telescopes are selected for this figure
(Sibyll/FLUKA simulations).
}
\end{figure}

The fraction of events which do not pass the 
threshold in the energy sum
E$_{\Sigma}(\pi{^0},\eta)$ can be used to accelerate the
simulation of background events for IACTs, using a two-pass approach.
In a first pass air showers are simulated without any Cherenkov photon
production. For CORSIKA this is about 20-100 times faster 
than simulations with Cherenkov photon production (depending
on the Cherenkov photon bunch size parameter). 
Next, only events which pass a certain
$\pi^{0}$ and $\eta$ energy sum threshold, are simulated with Cherenkov photon
production, accepting only a small loss in genuine $\gamma$-like proton showers.
The requirement of E$_{\Sigma}(\pi{^0},\eta) > 30$ GeV, for instance,
would reduce the number of fully simulated events by a factor of 9,
with a loss of less than 5\% of $\gamma$-like events (see Figure
\ref{fig-pi0-2b}).


Primaries other than protons have so far been neglected in this analysis,
although they compose about 25\% of all cosmic rays in this energy range.
Their contribution 
to $\gamma$-like showers is suppressed because the energy (E) of the primary 
is shared by several nucleons of energy E/A each. A single nucleon, even 
if it dumps all its energy into a $\pi^0$, can therefore only contribute 
E/A to the electromagnetic channel. To get the large fraction of E (say 80\%) 
required for a $\gamma$-like event, many of the nucleons would have to produce 
almost exclusively $\pi^0$s. The nucleons of a primary nucleus interact 
independently (the typical binding energy is much smaller than their energy).
The probability for a nucleon $p_n$ to produce predominantly 
electromagnetic secondaries (e.g.~$>$ 50\%) is already as low as 1\% (see 
Figure \ref{fig-pi0-3}, right panel). Therefore, the probability of A nucleons producing 
all predominantly electromagnetic output scales to first approximation with 
$(p_n)^A$, which is even for Helium of the order of $10^{-4}$.


\section{Influence of interaction models}
\label{chap-models}

Results from simulations depend in general on the choice of the
nuclear and hadronic interaction model. Models differ since those
interactions are not well known in the kinematical regions
relevant to cosmic rays (i.e.~high energies, very forward emission angles).
Therefore, models rely typically on phenomenological descriptions 
of interactions and extrapolate to the energy and angular
ranges required.
For the TeV range the extrapolation from collider experiments is still
moderate. Nevertheless, particle interactions (cross-sections, energy
spectra, multiplicity distributions, etc.) do differ significantly
for both low and high energy interaction models (see for example
\cite{Knapp-1996}, \cite{Knapp-1999}, \cite{Heck-2004}, \cite{Heck-2005}). 
The CORSIKA package allows the systematic study of
these differences since several low- and high-energy interaction
models are available in the same framework. 
The transition energy between high and low
interaction models varies from 50 GeV to 1 TeV, depending on the
model combination chosen.  
Here we examine the models 
FLUKA (version 2003.1)  \cite{Fasso-2005},
GHEISHA (version 2002) \cite{Fesefeldt-1985}, 
URQMD (version 1.3.1) \cite{Bass-1998} for low energies, and 
QGSJet (version 01c) \cite{Kalmykov-1997}
and Sibyll (version 2.1) \cite{Engel-1999}
for high-energy interactions.

\begin{figure}[p]
%
\centering\includegraphics[width=0.60\textwidth]{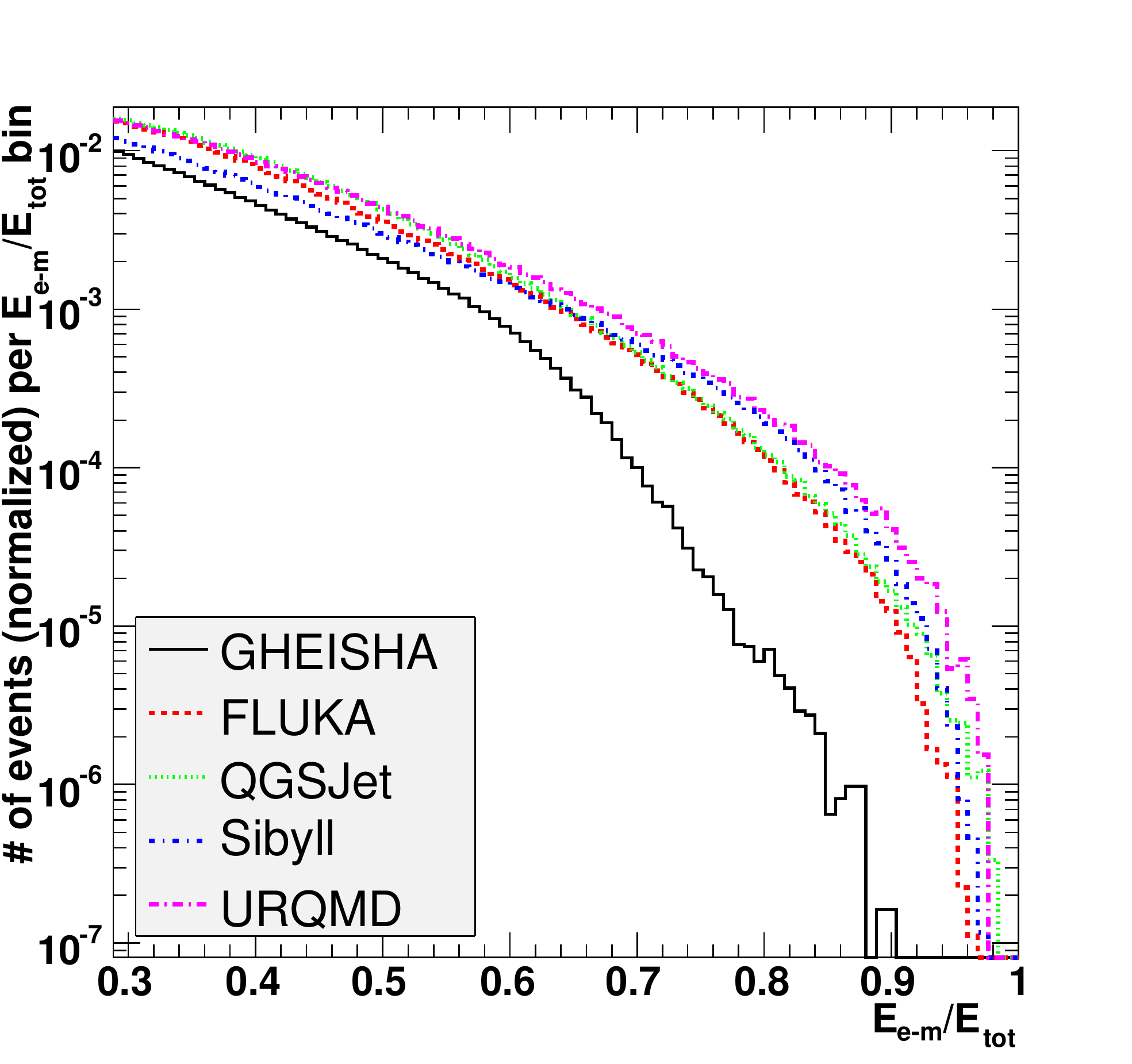}
%
\caption{\label{fig-fid-a}
Energy deposited in the electromagnetic components ($\gamma$, e${^\pm}$, $\pi^{0}$, $\eta$) in the
interactions of 100 GeV protons with nitrogen, for different interaction models.
The figure is normalized to the total number of simulated events.
}
\end{figure}

\begin{figure}[p]
\centering\includegraphics[width=0.65\textwidth]{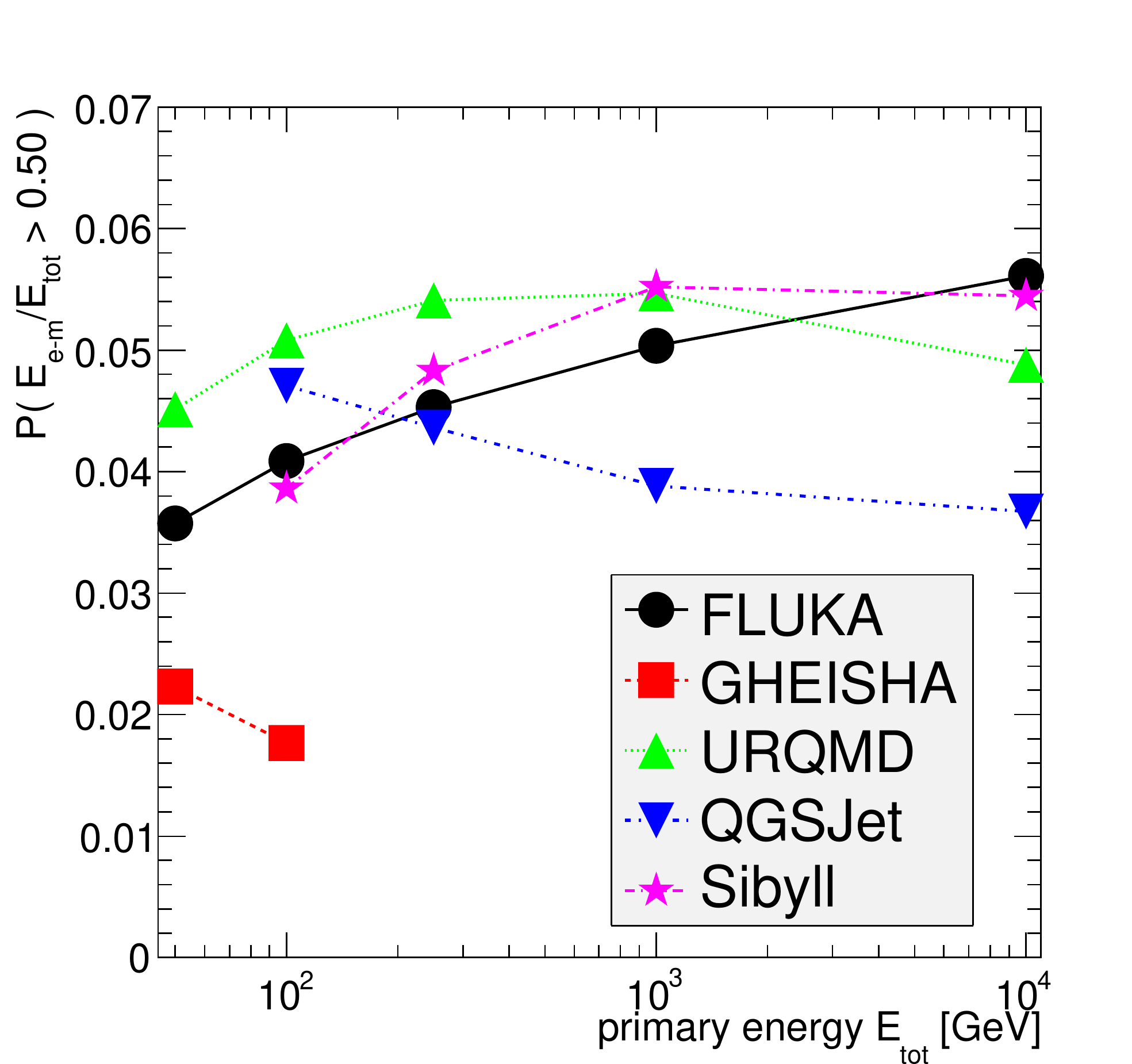}
\caption{\label{fig-fid-b}
Probability that 
more than 50\% of the primary energy is deposited
in the electromagnetic part in proton-nitrogen collisions 
for different interaction models and primary energies.
}
\end{figure}

Simulations of the interaction of protons with nitrogen nuclei
(i.e.~the first interaction in a proton-induced air shower) are used to study the
amount of energy deposited in the electromagnetic component right at the
start of the shower development.  Figure \ref{fig-fid-a}  shows
the distribution of the energy in the electromagnetic component for
100 GeV protons in different interaction models.  Figure \ref{fig-fid-b}
displays the probability that more than 50\% of the primary
energy is deposited in the electromagnetic component, as a function of
primary energy.  Note that not all models cover the whole energy
range.  QGSJet and Sibyll are not supposed to be used below about 80 GeV
primary energy, and GHEISHA is a low energy model, not suitable for energies
well above 100 GeV.

\begin{figure}[p]
\centering\includegraphics[width=0.65\textwidth]{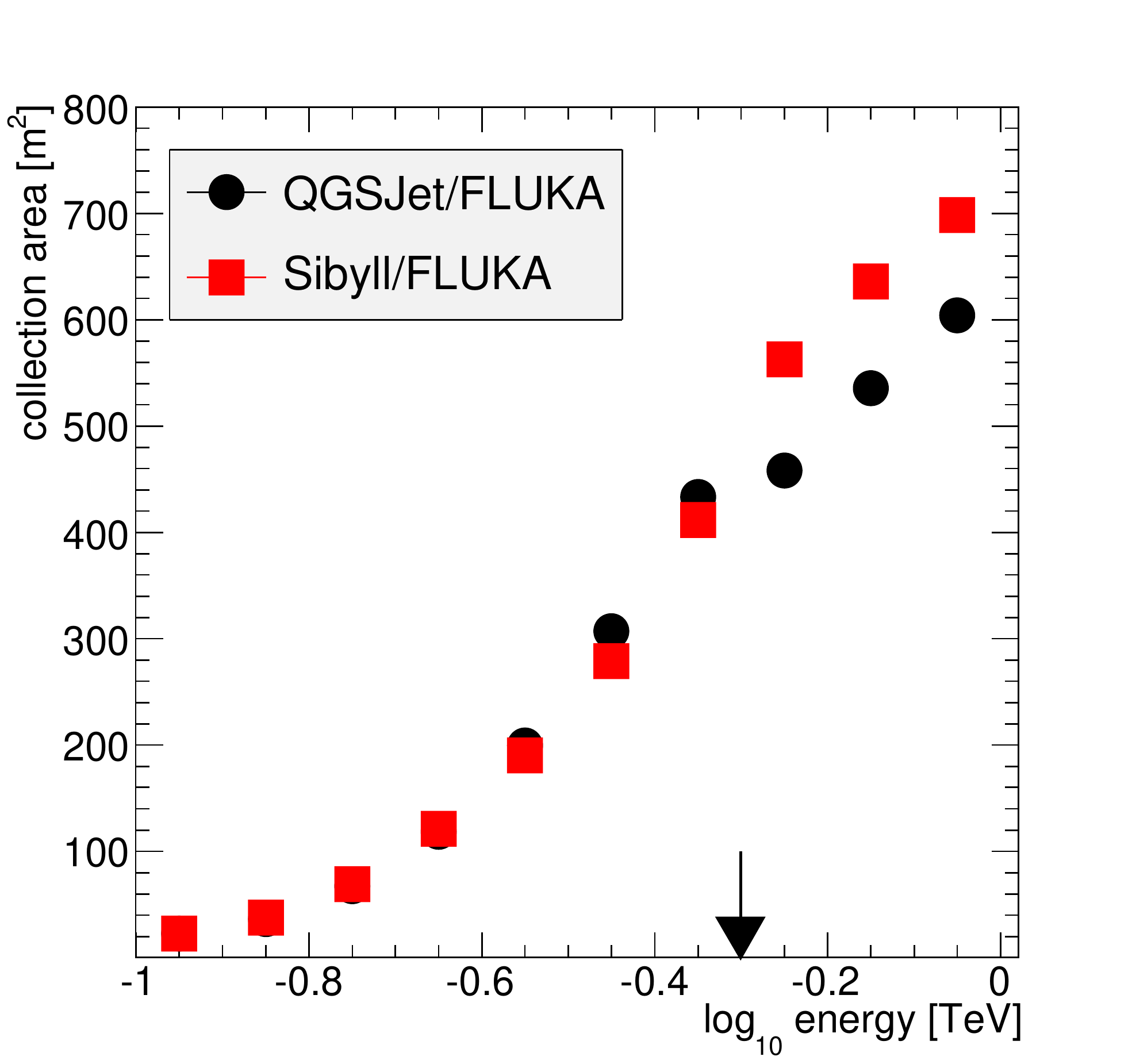}
\caption{\label{fig-effA}
Collection area of the considered array of four telescopes for proton showers for QGSJet/FLUKA
(transition energy 500 GeV) and Sibyll/FLUKA (transition energy 80 GeV) simulations.
The arrow indicates a primary energy of 500 GeV.
}
\end{figure}

GHEISHA gives a very different prediction compared to other
models: events with large $E_{e-m}/E_{tot}$ are less than half as
probable. According to ref.~\cite{Heck-2004}, GHEISHA does not
reproduce well the available experimental data of pion production and 
generates in general too few pions. 
Differences between the
other models are in the range of 20-40\%.  URQMD, Sibyll, and FLUKA
tend to deposit more energy in the electromagnetic part, QGSJet
systematically less.
However, in the past QGSJet was the most successful of the interaction 
models available, when compared with air shower data.
QGSJet employs the most sophisticated treatment of diffractive 
interactions and produces on average more secondaries than other
models.
Therefore, it is not surprising that QGSJet finds it more difficult to 
produce secondaries with large fractions of the primary energy. 

What is the effect of these differences on the Cherenkov 
photon part of the air shower and on the simulated performance of 
arrays of IACTs? Table \ref{tab-events} shows that 
the number of reconstructed and selected events of 
QGSJet/FLUKA and Sibyll/FLUKA simulations are very similar.
Also, the fraction of muonic $\gamma$-like events differ very little: 
however, these numbers are a convolution of 
the primary proton energy spectrum and the energy dependent
detection and selection efficiencies of the experiment.
Energy dependent measures like the collection area should give a
better description of the performance of IACTs.
The collection area of an instrument is an overall
measure of the number of Cherenkov photons in the shower, 
their angular and spatial distributions, 
and the energy dependence of it.
It is also affected by the probability and the shape of the images.

The energy dependent collection areas, after all $\gamma$-hadron separation cuts, of
QGSJet/FLUKA and Sibyll/FLUKA simulations are compared in Figure
\ref{fig-effA}. Note that the transition energy for the QGSJet/FLUKA
simulations is at 500 GeV, for Sibyll and FLUKA it is at 80 GeV
(and that also lower energy interactions play a role in the shower 
development of high energy showers).
The
collection areas are very similar in the energy range from 100 GeV to
500 GeV, where the figure shows essentially a comparison of pure FLUKA with
Sibyll/FLUKA simulations. However, the QGSJet/FLUKA collection area ($>$ 500 GeV)
shows a
discontinuity exactly at the transition energy
between QGSJet and FLUKA, with a smaller collection area at
higher energies. As shown in Figure \ref{fig-fid-b}, the difference in
the amount of energy between QGSJet, and FLUKA at 500 GeV deposited
into the electromagnetic part of the shower is about 15-30\%, while
the difference between Sibyll and FLUKA is below 5\%. This QGSJet
version, in contrast
to Sibyll and FLUKA, cannot reproduce the experimental values
of pion multiplicity in proton-proton interactions at energies of
about 500 GeV \cite{Heck-2004}.
A very similar analysis of events with a primary energy less than 100
GeV shows that the GHEISHA model predicts about 20\% less $\gamma$-like 
proton events with 3-fold array triggers. Interestingly,
the number of two-telescope events due to muons increases by about the
same amount.
The systematic differences of about 25\% in the predictions of
Sibyll/FLUKA and QGSJet/FLUKA at energies above 500 GeV for the
collection area translate directly into an uncertainty of about 10\%
for any sensitivity estimate.
QGSJet/FLUKA predicts a lower background, and therefore 
a higher sensitivity to $\gamma$-ray sources.
The choice of transition energies in this study
does not allow a definitive statement about the differences between QGSJet and 
Sibyll at energies below 500 GeV, but Figure \ref{fig-fid-b} indicates
that both models converge at about 200 GeV and give here results
similar to FLUKA.

Both findings, the different collection area of QGSJet at energies
above 500 GeV and the shortcomings of GHEISHA, indicate that a careful
choice of both, interaction models and transition energies, is necessary
to obtain reliable results.\footnote{During completion of this paper
a new version of QGSJet has been released, with marked improvements in the sub
TeV range.
\cite{Ostapchenko-2005}.}
 

%
%

\section{Summary}

With available simulation tools, air showers of primary $\gamma$-rays and cosmic rays, and
complex telescope systems can be modeled in sufficient detail to study the design and performance
of current and future imaging atmospheric Cherenkov telescope systems.

$\gamma$-ray shower simulations in the GeV - 100 TeV range have very small uncertainties, as the relevant physics is well known. 
 On the basis of these simulations $\gamma$-ray selection procedures are optimized 
and $\gamma$-induced showers can be securely identified. 

The dominant background, however, is due to cosmic rays, and some cosmic ray showers look 
very much like $\gamma$ showers. Simulations reveal which type of background events trigger the instruments and survive all $\gamma$-hadron selection cuts. They are mainly hadronic events which, by chance, transfer much of the primary's energy to electromagnetic subshowers during the first few interactions. Thus, interactions with low multiplicity and high $\pi^0$ fraction are the most important, and pose an irreducible background.
These showers are mainly produced by diffractive interactions with particle production in the very forward region, and because diffractive events are subject to considerable model uncertainties, the cosmic-ray background can only be estimated with about 20-40\% uncertainty, based on current models.

Events with energetic muons can produce $\gamma$-like images, even in stereoscopic systems, if the telescopes are close enough to be within the Cherenkov lightpool of the muon. However, a cut in the distance between the image centroids normalised by the telescope distance removes these events efficiently. 

The preselection of showers with a high energy fraction in electromagnetic particles can be used to reduce overall computing time for background calculations by approximately an order of magnitude. 

\section*{Acknowledgments}

G.M. acknowledges the support as a Feodor Lynen Fellow of the Alexander von Humboldt foundation. 
He would also like to thank the High Energy Astrophysics Group of the University of Leeds for
hospitality during the main phase of this work.


\end{document}